%% file: ms.tex
\documentclass[preprint]{sigplanconf}
\usepackage[T1]{fontenc}
\usepackage[utf8]{inputenc}

\usepackage{flushend}
\usepackage{amsmath}
\usepackage{amsthm}

\usepackage{graphicx}
\usepackage{float}
\usepackage{amsmath}
\usepackage{amssymb}
\usepackage{listings}
\usepackage{hyperref}
\usepackage{multirow}
\usepackage{tabularx} 
\usepackage{fancyvrb} 
\usepackage{tikz} 
\usepackage{pgfplots}
\pgfplotsset{compat=newest} 
\usetikzlibrary{patterns}
\floatstyle{ruled}
\newfloat{algorithm}{tbp}{loa}
\providecommand{\algorithmname}{Listing}
\floatname{algorithm}{\protect\algorithmname}

\begin{document}

\setlength{\pdfpageheight}{\paperheight}
\setlength{\pdfpagewidth}{\paperwidth}



\titlebanner{~}        
\preprintfooter{~}   

\title{Towards Automatic Transformation of  Legacy Scientific Code into OpenCL for Optimal Performance on FPGAs}

\authorinfo{Wim Vanderbauwhede \and Syed Waqar Nabi}
           {School of Computing Science\\University of Glasgow, Glasgow, UK}
           {\{wim.vanderbauwhede,syed.nabi\}@glasgow.ac.uk}

\maketitle

\begin{abstract}
There is a large body of legacy scientific code written in languages like Fortran that is not optimised to get the best performance out of heterogeneous acceleration devices like GPUs and FPGAs, and manually porting such code into parallel languages frameworks like OpenCL requires considerable effort. We are working towards developing a turn-key, self-optimising compiler for accelerating scientific applications, that can automatically transform legacy code into a solution for heterogeneous targets. In this paper we focus on FPGAs as the acceleration devices, and carry out our discussion in the context of the OpenCL programming framework. We show a route to automatic creation of kernels which are optimised for execution in a \emph{streaming} fashion, which gives optimal performance on FPGAs. We use a 2D shallow-water model as an illustration; specifically we show how the use of \emph{channels} to communicate directly between peer kernels and the use of on-chip memory to create stencil buffers can lead to significant performance improvements. Our results show better FPGA performance against a baseline CPU implementation, and better energy-efficiency against both CPU and GPU implementations.
\end{abstract}



\keywords
FPGA, GPU, Accelerators, OpenCL, Scientific Computing, High-Performance Computing, Compiler, Heterogeneous Computing.

%
%
%
%
%
%
%
%
%
%
%
%

\section{Introduction}
\subsection{Automatic acceleration of legacy scientific code}

\begin{figure}
	\centering{}\includegraphics[width=1.0\linewidth]{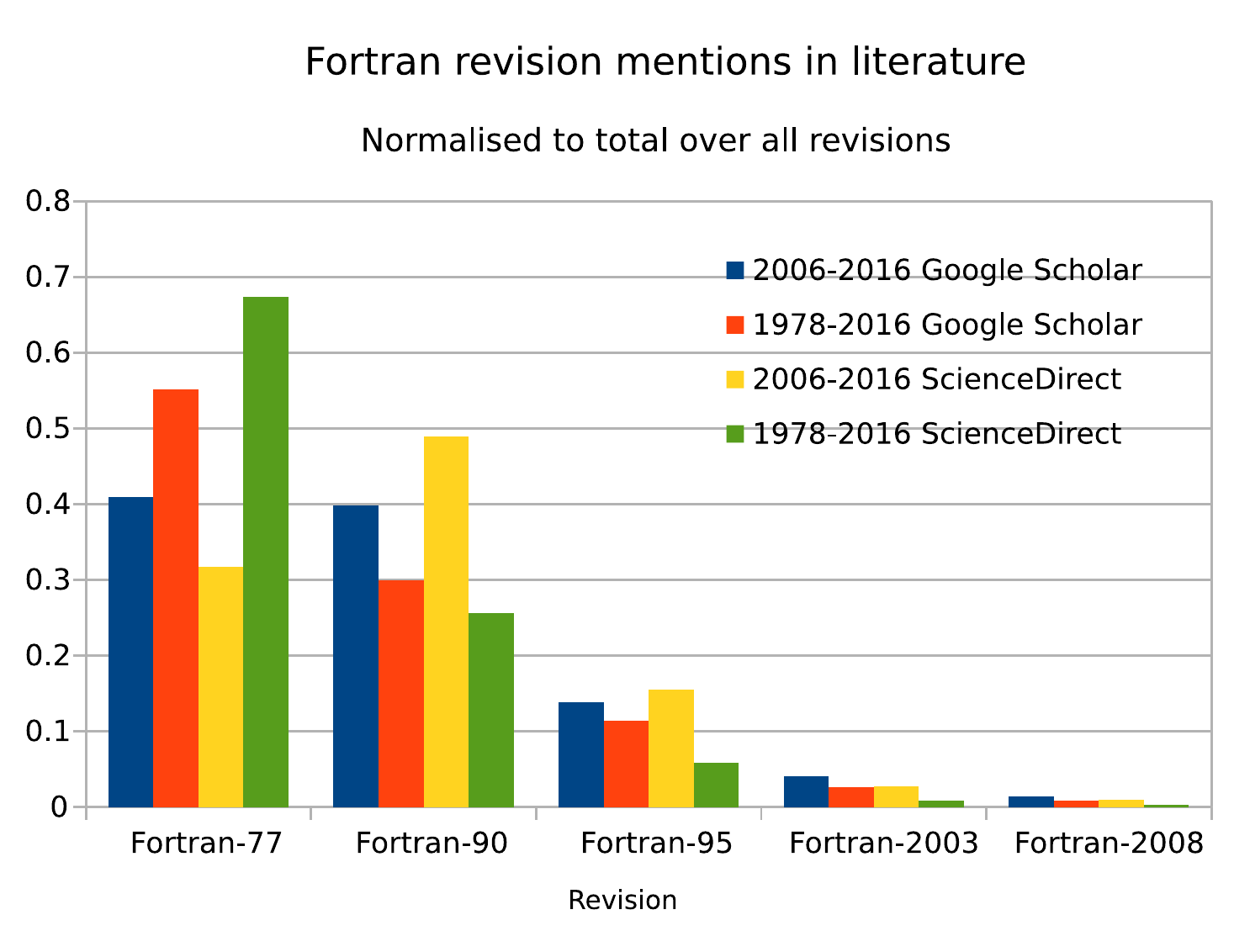}\caption{\label{fig:Literature-mentions-of}Literature mentions of different
		revisions of Fortran using Google Scholar and ScienceDirect}
\end{figure}

A considerable proportion of scientific code in use --  ``legacy'' and new -- is still effectively written in FORTRAN 77. A comparison of relative citations for each of the main revisions of Fortran  (from Google Scholar and ScienceDirect) is shown in \autoref{fig:Literature-mentions-of}%
\footnote{Citations per revision are normalized to sum of citations for all revisions.}
. The results have been collected for
the past 10 years (2006-2016), and additionally since the release of FORTRAN
77 (1978-2016). For an absolute reference, note that there were 15,700 Google Scholar citations
mentioning FORTRAN 77 (2006--2016). It
is obvious that FORTRAN 77 is still widely used, and that newer 
standards (2003, 2008) have not yet penetrated enough for widespread adoption.

We can see from the above evidence -- and our own anecdotal evidence from collaboration with scientists confirms this -- that FORTRAN 77 is still the language of choice for writing scientific models. This is in addition to the vast amount of  FORTRAN 77 legacy
code already in existence. Since the FORTRAN 77 language was designed with very different 
requirements and assumptions compared to today's, code written
in it has inherent issues with readability, scalability, maintainability
and parallelization. A comprehensive discussion of the issues can
be found in \cite{tinetti2012fortran}. As a result, there have been considerable efforts
at refactoring legacy code, either interactive or
automatic, and to address one or several of these issues. 

This work is in some ways part of that effort, but we are specifically interested
in automatically refactoring Fortran for OpenCL-based accelerators in general, and FPGAs in particular.
In this paper we present a source-source compilation approach to transform
sequential FORTRAN 77 legacy code into high-performance OpenCL-accelerated
programs with auto-parallelized kernels with a focus on FPGA implementation, without need for directives
or extra information from the user.

\subsection{Accelerating Scientific Code using FPGAs}

There is a strong motivation to target FPGAs as acceleration devices 
. The domain of target devices and machines for high-performance scientific computing has become increasingly heterogeneous in recent years. There is a general consensus that no single type of device -- CPU, GPU or FPGA -- will be optimal across the range of scientific applications. GPUs have established themselves as a potent alternative to conventional CPUs for accelerating scientific applications.  That a large number of supercomputers in the top 500 list contain GPU accelerators is an indication of this trend. FPGAs are a more recent addition on this canvas of high-performance computing (HPC), and in spite of significant improvements in recent years, they are still from far from widespread adoption as mainstream acceleration devices. 

A key challenge that applies in a lesser or greater extent to all kinds of accelerators is to write parallel, high-performance code that is optimized for performances on that device. The challenge is all the more acute for FPGAs, which are notoriously difficult to program. Improvements in FPGA logic capacity as well as high-level synthesis (HLS) programming frameworks like Altera's AOCL, Xilinx's Sdaccel, and Maxeler, have begun the FPGA transition from peripheral or embedded devices to first-order desktop and server computing devices. However, FPGAs have failed to make the kind of inroads that GPUs have made in the last decade. This is, in part at least, due to the fact that until very recently there were no practical high-level programming platforms like OpenCL for FPGAs, and even with their introduction it is challenging to write high-performance code. It is our contention that such challenges will remain a hurdle to the adoption of acceleration devices -- especially FPGAs -- in mainstream HPC, and that high level programming frameworks like OpenCL should themselves be targets for still higher level compilers that can work with sequential, unoptimized legacy code.



\begin{figure}
	\centering
	\includegraphics[width=1.0\linewidth]{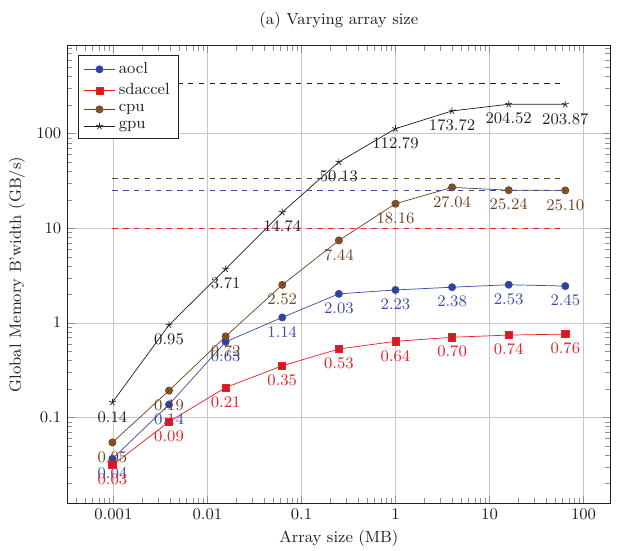}
	\caption{Bandwidth to contiguous data in the external/main memory for four different targets while varying (a) array sizes and (b) vector size (memory coalescing). Two of the targets (AOCL and Sdaccel) are FPGAs (Stratix-V and Virtex-7).}
	\label{fig:mpstream}
\end{figure}

FPGAs have a fundamentally different architecture from CPUs and GPUs, and are optimized for performance in their own unique way, even if the code is written in the same heterogeneous parallel programming language like OpenCL. FPGAs consist of a fine-grained reconfigurable fabric that can be synthesized into a custom circuit for a given application. 
They are relatively low frequency devices, and when competing against other high-performance devices such as GPUs, the key for FPGAs is to create custom, deep pipelines for given application kernels, and maintain maximum continuity and minimum disruption in the resulting dataflow stream. 
Maintaining a continuous stream of data on the FPGA pipeline without disrupting it with repeated accesses to its external DRAM memory (\emph{global memory} in OpenCL terminology) is also important because of the relatively low access bandwidth, especially for random access. In previous work \cite{nabi-mpstream-h2rc}, we have adapted a synthetic benchmark to show the relative bandwidth to global memory of various devices, and \autoref{fig:mpstream} highlights the magnitude of discrepancy between GPUs and FPGAs. 

\section{Related Work}

\subsection{FPGA acceleration of scientific code}
There are a number of commercial tools available that provide a high-level programming route for accelerating scientific code on FPGAs. Maxeler \cite{112.154} is a good example of such an HLS design framework. It provides a Java meta-programming model for describing computation
kernels and connecting data streams between them. It has been used for accelerating applications from various scientific and financial domains. 
Altera OpenCL or AOCL \cite{112.151} is  the Altera (now Intel) implementation of the OpenCL heterogeneous parallel programming framework for their FPGAs. While it is based on the OpenCL standard, it has vendor specific optimization extensions, one of which is the use of \emph{channels} to communicate directly between kernels%
\footnote{Altera introduced these channels before the OpenCL standard adopted \emph{pipes} which is more or less the same thing. Now that the OpenCL standard has pipes, one can use either channels or pipes in AOCL.}
. 
Xilinx similarly has its own OpenCL implementation called SDAccel \cite{sdaccel}, which is based on its more mature Vivado HLS tool\cite{vivado}. Like AOCL, SDAccel is based on the OpenCL standard, and also  proposes custom optimizations to improve performance.
The OpenCL-FPGA tools like AOCL and SDAccel have been shown to get good performance and energy-efficiency, e.g. \cite{Weller:2017:EES:3020078.3021730} demonstrates their use for PDE applications in the context of power-constrained applications. \cite{7051218} carried out a suitability analysis of FPGAs for heterogeneous HPC platforms, using the Berkeley 13 dwarfs as a reference. They found FPGAs to be suitable for 5 of the 13 dwarves, but noted that they are difficult to use for non-specialized designers, and emphasized the importance of more abstraction and less customization. 

While these and other similar works have significantly advanced the state of the art in the use of FPGAs for scientific computing using high-level programming, the current tools and workflows still require one to manually write code in parallel languages like MaxJ (for Maxeler machines) or OpenCL, and optimize it for FPGAs, both of which are significant tasks. Our tool creates OpenCL code automatically by refactoring Fortran code, and we show a route to optimizing this OpenCL code for FPGAs.

\subsection{Refactoring of legacy code}


There are a number of source-to-source compilers and refactoring tools
for Fortran available. However, very few of them actually support
FORTRAN 77. The most well known are the ROSE framework\footnote{\url{http://www.rosecompiler.org/index.html}}
from LLNL \cite{liao2010rose}, which relies on the Open Fortran Parser
(OFP)\footnote{\url{http://fortran-parser.sourceforge.net/}} . This
parser claims to support the Fortran 2008 standard. Furthermore, there
is the language-fortran\footnote{\url{https://hackage.haskell.org/package/language-fortran}}
parser which claims to support FORTRAN 77 to Fortran 2003. A  refactoring
framework which claims to support FORTRAN 77 is CamFort \cite{Orchard:2013:UFS:2541348.2541356},
according to its documentation it supports Fortran 66, 77, and 90
with various legacy extensions.


Like CamFort, the Eclipse-based interactive refactoring tool Photran
\cite{overbey2005refactorings}, which supports FORTRAN 77 - 2008,
is not a whole-source compiler, but works on a per-file basis (which
is in fact what most compilers do). Both CamFort and Photran provide
very useful refactorings, but these are limited to the scope of a
code unit. For effective refactoring of common blocks, and determination
of data movement direction, as well as for effective acceleration,
whole-source code (inter-procedural) analysis and refactoring is essential. 

A long-running project which does support inter-procedural analysis
is PIPS\footnote{\url{http://pips4u.org/}}, started in the 1990's.
The PIPS tool does support FORTRAN 77 but does not supported the refactorings
we propose. Support for auto-parallelization via OpenCL was promised
\cite{amini2011pips} but has not yet materialized. None of the other refactoring tools support either auto-parallelization or OpenCL.

\section{Heterogeneous Computing for Accelerating Scientific Applications}
\label{subsec:Heterogenenous-Computing-and}
Many scientific codes
have already been investigated for and ported manually to GPUs as well as FPGAs, and
excellent performance benefits have been reported. There are many
approaches to programming accelerators, but we restrict our discussion
to open standards and to solutions that work in Fortran.

\subsection{OpenCL}

The OpenCL framework\cite{stone2010opencl} presents an abstraction
of the accelerator hardware based on the concept of \emph{host} and
\emph{device}. A programmer writes one or more \emph{kernels} that
are run directly by the accelerator and a \emph{host program} that
is run on the system's main CPU. The host program handles memory transfers
to the device and initializing computations and the kernels do the
bulk of the processing, in parallel on the device. In general OpenCL assumes separate memory spaces for the host and device and assumes a shared-memory programming mode for the device kernels.
The main advantage
of OpenCL over proprietary solutions such as e.g. CUDA (to which it
is very similar) is that it supported by a wide range of devices,
including multicore CPUs, FPGAs and GPUs. From the programmer perspective,
OpenCL is very flexible but quite low level and requires a lot of
boilerplate code to be written. This is a considerable barrier for
adoption by scientists. Furthermore, there is no official Fortran
support for OpenCL: the host API is C/C++, the kernel language is
based on a subset of C99. To remedy this we have developed \cite{vanderbauwhede2015twinned}
a Fortran API for OpenCL\footnote{https://github.com/wimvanderbauwhede/OpenCLIntegration}.

\subsection{OpenACC and OpenMP}

OpenACC\footnote{https://www.openacc.org/} takes a directive based
approach to heterogeneous programming that affords a higher level
of abstraction for parallel programming than OpenCL or CUDA. In a
basic example, a programmer adds \emph{pragmas} (compiler directives)
to the original (sequential) code to indicate which parts of the code
are to be accelerated. The new source code, including directives,
is then processed by the OpenACC compiler and programs that can run
on accelerators are produced. There are a number of extra directives
that allow for optimization and tuning to allow for the best possible
performance.

With OpenMP version 4, the popular OpenMP standard\footnote{http://www.openmp.org/}
for shared-memory parallel programming now also supports accelerators.
The focus of both standards is slightly different, the main difference
being that OpenMP allows conventional OpenMP directives to be combined
with accelerator directives, whereas OpenACC directives are specifically
designed for offloading computation to accelerators.

Both these annotation-based approaches are local: they deal with parallelization
of relatively small blocks and are not aware of the whole code base,
and this makes them both harder to use and less efficient. To use
either on legacy FORTRAN 77 code, it is not enough to insert the pragmas:
the programmer has to ensure that the code to be offloaded is free
of global variables, which means complete removal of all common block
variables or providing a list of shared variables as annotation. The
programmer must also think carefully about the data movement between
the host and the device, otherwise performance is poor. 

\subsection{Raising the abstraction level}

Our approach allows an even higher level of abstraction than that
offered by OpenACC or OpenMP: the programmer does not need to consider
how to achieve program parallelization, but only to mark -- using a
single annotation -- which subroutines will be paralleled and offloaded
to the accelerator. Our compiler provides a fully automatic conversion
of a complete FORTRAN 77 codebase to Fortran 95 with OpenCL kernels.
Consequently, the scientists can keep writing the code in FORTRAN
77, and the original code base remains always intact.

\section{A Refactoring Source-to-Source Compiler for FORTRAN 77}
FORTRAN 77 code is often computationally efficient, and programmer
efficient in terms of allowing the programmer to quickly write code
and not be too strict about it. As a result it becomes quickly very difficult
to for maintain and port. Our goal is  that the refactored code should
meet the following requirements:

\subsection{Modern, Maintainable and Extensible }

FORTRAN 77 was designed with very different requirements from today's
languages, notably in terms of avoiding bugs. It is said that C gives
you enough rope to hang yourself. If that is so then FORTRAN 77 provides
the scaffold as well. Specific features that are unacceptable in a
modern language are:
\begin{itemize}
\item Implicit typing, i.e. an undeclared variable gets a type based on
its starting letter. This may be very convenient for the programmer
but makes the program very hard to debug and maintain. Our compiler
makes all types explicit (\texttt{implicit none}).
\item No indication of the intended access of subroutine arguments: in FORTRAN
77 it is not possible to tell if an argument will be used read-only,
write-only or read-write. This is again problematic for debugging
and maintenance of code. Our compiler infers the \texttt{intent} for
all subroutine and function arguments. 
\item In FORTRAN 77, procedures defined in a different source file are not
identified as such. For extensibility as well as for maintainability,
a module system is essential. Our compiler converts all non-program
code units into modules which are \texttt{use}d with an explicit export
(\texttt{only}) declaration. 
\end{itemize}
There are several more refactorings that our compiler applies, such
as rewriting label-bases loops as do-loops etc., but they are less
important for this paper.

\subsection{Accelerator-ready}

As mentioned in 
\autoref{subsec:Heterogenenous-Computing-and},
the common feature of the vast majority of current accelerators is
that they have a separate memory space, usually physically separate
from the host memory. Furthermore, the common offload model is to
create a ``kernel'' subroutine (either explicitly or implicitly)
which is run on the accelerator device. Consequently, it is crucial
to separate the memory spaces of the kernel and the host program. 
\begin{itemize}
\item FORTRAN 77 programs makes liberal use of global variables through
``\texttt{common}'' blocks. Our compiler converts these common block
variables into subroutine arguments across the \emph{complete} call
tree of the program. Although refactoring of common blocks has been
reported for some of the other projects, to our knowledge our compiler
is the first to perform this refactoring across multiple nested procedure
calls, potentially in different source code units. 
\end{itemize}

\subsection{Code Transformation Validation\label{sec:Validation}}

To assess the correctness and capability of our refactoring compiler,
we used the NIST (US National Institute of Standards and Technology)
FORTRAN 78 test suite \footnote{http://www.itl.nist.gov/div897/ctg/fortran\_form.htm},
which aims to validate adherence to the ANSI X3.9-1978 (FORTRAN 77)
standard. We used a version with some minor changes\footnote{http://www.fortran-2000.com/ArnaudRecipes/fcvs21\_f95.html}
This test
suite comprises about three thousand tests organized into 192 files.
We skipped a number of tests because they test features that our compiler
does not support. In particular, we skipped tests that use spaces
in variable names and keywords (3 files, 23 tests) and tests for corner
cases of common blocks and block data (2 files, 37+16 tests). After
skipping these types of tests, 2867 tests remain, in total 187 files
for which refactored code is generated. The test bench driver provided
in the archive skips another 8 tests because they relate to features
deleted in Fortran 95. In total the test suite contains 72,473 lines
of code (excluding comments). Two test files contain tests that fail
in gfortran 4.9 (3 tests in total).

Our compiler successfully generates refactored code for all tests,
and the refactored code compiles correctly and passes all tests (2864
tests in total).

Furthermore, we tested the compiler on a simple 2-D shallow water
model from \cite{kampf2009ocean} (188 loc) and on four real-word
simulation models: the Large Eddy Simulator for Urban Flows \footnote{https://github.com/wimvanderbauwhede/LES},
a high-resolution turbulent flow model\cite{vanderbauwhede2015twinned}
(1,391 loc); the shallow water component of Gmodel\footnote{http://www.sciamachy-validation.org/research/CKO/gmodel.html},
an ocean model\cite{burgers2002balanced} (1,533 loc); Flexpart-WRF\footnote{https://github.com/sajinh/flx\_wrf2},
a version of the Flexpart particle dispersion simulator\cite{brioude2013lagrangian}
that takes input data from WRF (13,829 loc); and the Linear Baroclinic
Model\footnote{http://ccsr.aori.u-tokyo.ac.jp/\textasciitilde{}hiro/sub/lbm.html},
an atmospheric climate model\cite{watanabe2003moist} (39,336 loc). 

Each of these models has a different coding style, specifically in
terms of the use of common blocks, include files, etc that affect
the refactoring process. All of these codes are refactored fully automatically
without changes to the original code and build and run correctly.
The performance of the original and refactored code is the same in
all cases.

\section{Auto-Parallelization in OpenCL}
\label{sec:auto-parallel-opencl}

Our goal is to convert legacy FORTRAN 77 code into parallel
code so that the computation can be accelerated using OpenCL. We
use a three-step process:

First, the above refactorings\footnote{https://github.com/wimvanderbauwhede/RefactorF4ACC}
result in a modern, maintainable, extensible and accelerator-ready
Fortran 95 codebase. This is an excellent starting point for many
of the other existing tools, for example the generated code can now
easily be paralleled using OpenMP or OpenACC annotations, or further
refactored if required. However, we want
to provide the user with an end-to-end solution that does not require
any annotations. 

The second step in our process is to identify data-level parallelism
present in the code in the form of \emph{maps} and \emph{folds} (i.e.
loops without dependencies and reductions). The terms \emph{map} and
\emph{fold} are taken from functional programming and refer to ways
of performing a given operation on all elements of a list. Broadly
speaking these constructs are equivalent to loop nests with and without
dependencies, and as Fortran is loop-based, our analysis in indeed
an analysis of loops and dependencies. However, our internal representation
uses the functional programming model where \emph{map} and \emph{fold}
are functions operating on other functions (i.e. they are higher-order
functions), the latter being extracted from the bodies of the loops.
Thus we raise the abstraction level of our representation and make
it independent of both the original code and the final code to be
generated. We apply a number of rewrite rules for map- and fold-based
functional programs (broadly speaking equivalent to loop fusion or
fission) to optimist the code.

The third step is to generate OpenCL host and device code from the
paralleled code. Because of the high abstraction level of our internal
representation, we could easily generate OpenMP or OpenACC annotations,
CUDA or Maxeler's MaxJ language used to program FPGAs. Our compiler\footnote{https://github.com/wimvanderbauwhede/AutoParallel-Fortran}
also minimizes the data transfer between the host and the accelerator
by eliminating redundant transfers. This includes determining which
transfers need to be made only once in the run of the program.

\section{Porting OpenCL Applications for Performance on FPGAs}
\label{sec:porting-for-perf-on-FPGAs}
In this section we discuss our ongoing work of extending the auto-parallelizing framework previously presented to generate OpenCL that is customized for performance on FPGAs. 

The potential to get good performance and energy efficiency on FPGAs is widely recognized but coupled with the realization that achieving the potential is not straightforward \cite{7051218}. Even if we use a \emph{code portable} framework like OpenCL, code written for CPUs or GPUs will not be \emph{performance portable} to FPGAs, and will require considerable optimization effort. In our view, some important guidelines for creating architectures on FPGAs that give high throughput are:

\begin{enumerate}
	\item Create deep and custom pipelines that make optimal use of on-chip resources.
	\item Maintain streaming on these pipelines.
	\item Minimize random access to external memory.
	\item Use (scarce) on-chip memory resources efficiently
	\item Use vendor-optimizations where suitable.
	\item Use optimized numerics where possible.
\end{enumerate}

In our view, the most critical requirement is to maintain streaming on a deep, custom pipeline, which relates to the first four points in the previous list, and is the focus of our current work. The importance of this requirement of maintaining streaming is reflected e.g. in the Maxeler HLS framework, where an ``Open Space Programming'' paradigm is proposed\cite{openspl}: all variables in kernel code are by default dataflow variables, and not the content of a particular address(es) in a memory space. The OpenCL framework cannot adopt this dataflow paradigm at a fundamental level like Maxeler as it is a heterogeneous framework primarily driven by the GPU community. However, FPGA vendors like Altera and Xilinx who have adopted OpenCL as their HLS framework use numerous automatic and manual optimizations in order to create a streaming, pipelined, dataflow architecture. This is indicated by the adoption of custom ``channels'' by AOCL before the functionally similar ``pipes'' were officially made part of the OpenCL standard. These channels allow one to create coarse-grained pipelines across kernels, minimizing accesses to external memory. Also important is to carefully implement computations which involve stencils -- frequently encountered in scientific models -- as they can lead to random and redundant accesses to memory, killing the performance.

In \autoref{sec:auto-parallel-opencl}, we have already discussed our auto-parallelizing refactoring tool that can generate OpenCL code from Fortran 77 code. We propose to extend this tool to optimize the generated OpenCL code for FPGA implementation, and focus on two particular optimizations that encourage streaming and minimize random accesses to global memory, leading to improved performance: 1) Using FIFO \emph{channels} (or \emph{pipes}) for direct communication between kernels instead of communication via global memory and 2) avoiding random and redundant accesses to global memory for stencil computations by using FPGA on-chip buffers. These approaches are best illustrated through a real-world example, as discussed in the next section. 

\section{Illustration: 2D Shallow Water Model}
To illustrate our approach for extending our auto-parallelizing tool for FPGAs, we use a simple but realistic scientific model from the domain of computational fluid dynamics (CFD), called the \emph{shallow water} model. The model can be used to simulate waves where the horizontal scale is far greater than the depth of the body of water, so it can be applied for both shallow and deep portions of the ocean or the atmosphere. We use the shallow-water model described in \cite{kampf2009ocean} along with the provided Fortran model as our starting point. 

The model assumes wave periods that are short compared to the inertial period, so we can ignore the Coriolis force, and frictional effects along with other non-linear terms are ignored. With these simplifications, the equations governing the dynamics are reduced to:

\begin{equation*}
\frac{\partial u}{\partial t}=-g\frac{\partial \eta}{\partial x}
\end{equation*}

\begin{equation*}
\frac{\partial v}{\partial t}=-g\frac{\partial \eta}{\partial y}
\end{equation*}

\begin{equation*}
\frac{\partial \eta}{\partial t}=-\frac{\partial (uh)}{\partial x}-\frac{\partial (vh)}{\partial y}
\end{equation*}

where $u$ and $v$ are the two components of the horizontal velocity, $h$ is the total water depth, and $\eta$ is the sea-level elevation (so $h=h_{0}+\eta)$. 

The numerical simulation of these equations takes the form of the following three functions or \emph{kernels}, which are run across the 2D grid inside time loop.

\begin{description}
	\item[Dynamics] This kernel implements the dynamics of the model as described in the expressions shown above, computed as finite-difference (FD) equations. 
	\item[Filter] One of the artefacts of using FD methods is the introduction of artificial, numerical waves, and a first order \emph{Shapiro} filter is used to remove those.
	\item[Update] This kernel simply updates the height $h$ based on the elevation $\eta$ calculated for the current time step.
\end{description}

The model uses single-precision floats, and the original implementation is in Fortran-95\footnote{The code has artefacts of Fortran-77 coding style, so it may be fairer to call it a mixture of Fortran 77  and 95.}.

\subsection{Profiling the Fortran Code}
\begin{figure}[!ht]
	\centering
	\includegraphics[width=0.5\linewidth]{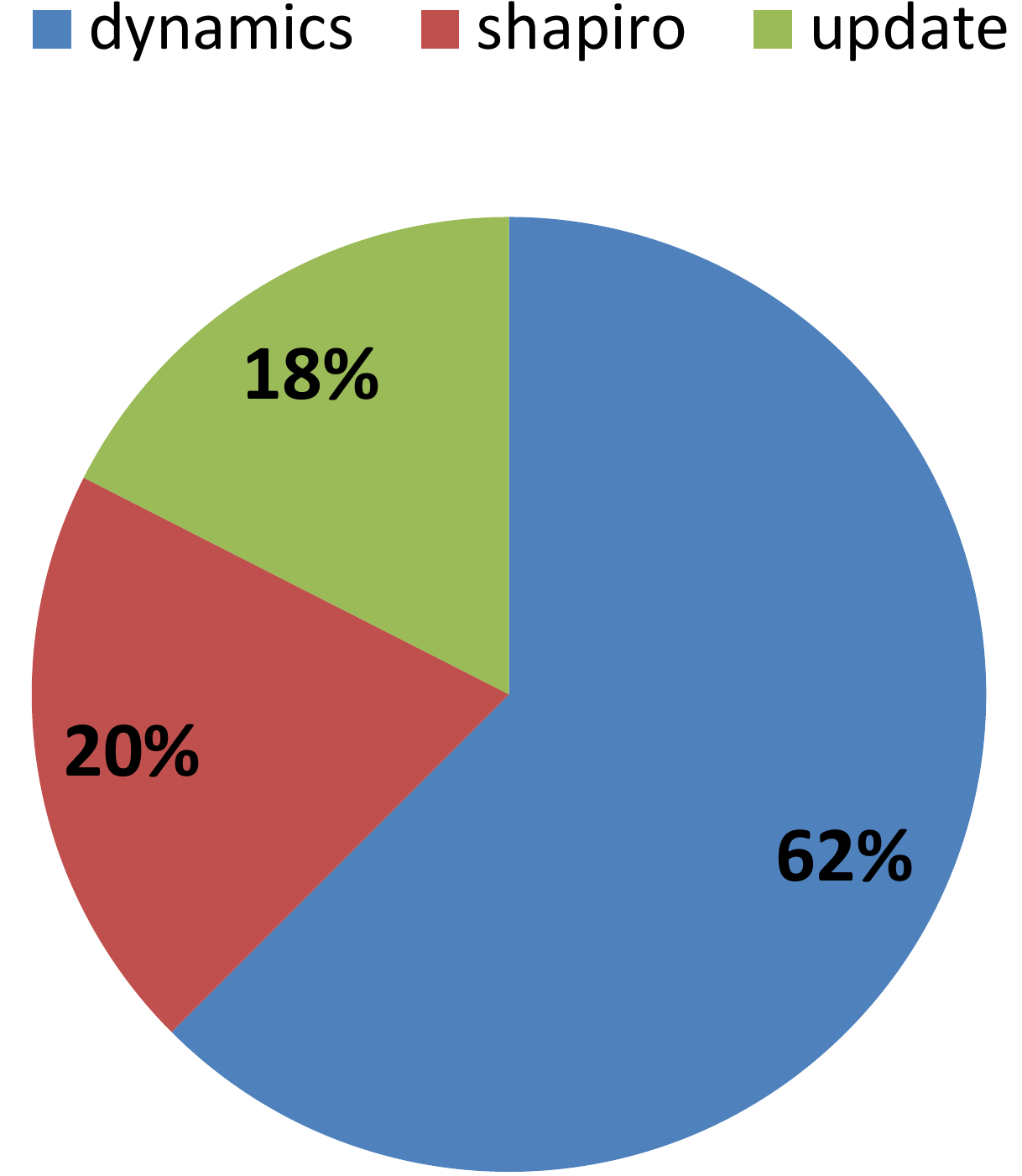}
	\caption{Dynamic profile of the 2D Shallow Water Code.}
	\label{fig:profile}
\end{figure}

We built a dynamic profile of the code using \emph{gprof}, and as one would expect, modelling the dynamics takes up most of the simulation (wall-clock) time. \autoref{fig:profile} shows the distribution of the simulation time across the three kernels, and while \emph{dynamics} take the most time, we can see that performance dividends can be expected by accelerating all three of the kernels on an accelerator. 

\subsection{Parallelization of the Code for GPU Implementations}
The auto-parallelizer tool described in \autoref{sec:auto-parallel-opencl} was used to generate OpenCL for GPU implementation. Our compiler automatically transforms the original code into three map-style kernels.


The results shown in \autoref{fig:results-2d-shallow-waterperf}
for domain size
of 500x500 and 10,000 time steps. This is
a high-resolution simulation with spatial resolution of 1 m and a
time step of 0.01 s. The automatically generated code running on GPU
is up to 9x faster than the original code. This is the same performance
as obtained by manual porting of the code to OpenCL.

\subsection{Parallelization of the Code for FPGA Implementations}
\label{sec:parallelize-for-fpgas}
As discussed in \autoref{sec:porting-for-perf-on-FPGAs}, our view is that the most important optimizations are those that allow us to maintain streaming on the device, which include a well known technique of introducing channels between the peer kernels, and our own innovation of using ``smart caches'' to optimize stencil computations. 

\subsubsection{The Baseline Version}
\begin{figure}[t!]
\centering
\includegraphics[width=1.0\linewidth]{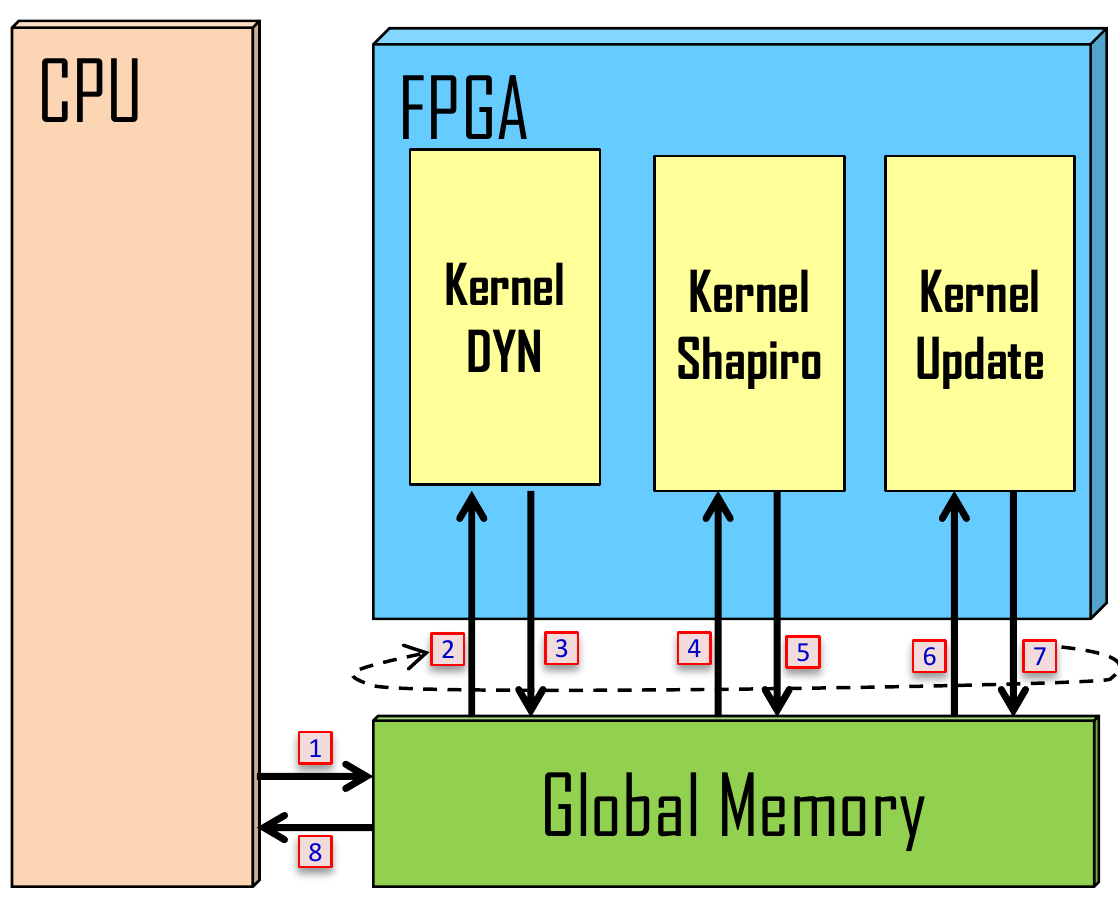}
\caption{The baseline FPGA implementation of the 2D shallow-water model on a Stratix-V FPGA using the Altera-OpenCL framework. The numbers show the sequence of operations. Note that steps 2--7 repeat $NT$ times, where $NT$ is the number of time steps of the simulation.}
\label{fig:2dshallow_FPGA_naive}
\end{figure}

If we naively take the OpenCL code for CPUs or GPUs, and port it to FPGA as-is, we get the baseline FPGA architecture as shown in \autoref{fig:2dshallow_FPGA_naive}.

The host transfers all the arrays to the device global memory (i.e., the external DRAM on the FPGA board) at the beginning of the execution. Then each of the three kernels on the device reads the relevant data from the global memory, processes it, and writes it back, one after the other. The kernels are called in a time loop $NT$ times, which in our experiments is $10,000$ steps. The kernels are communicating via the global memory, and it is immediately obvious that we are ferrying data back and forth many times between the FPGA and the global memory redundantly. 
 
\subsubsection{The Channelized Version}
\begin{figure}[t!]
\centering
\includegraphics[width=1.0\linewidth]{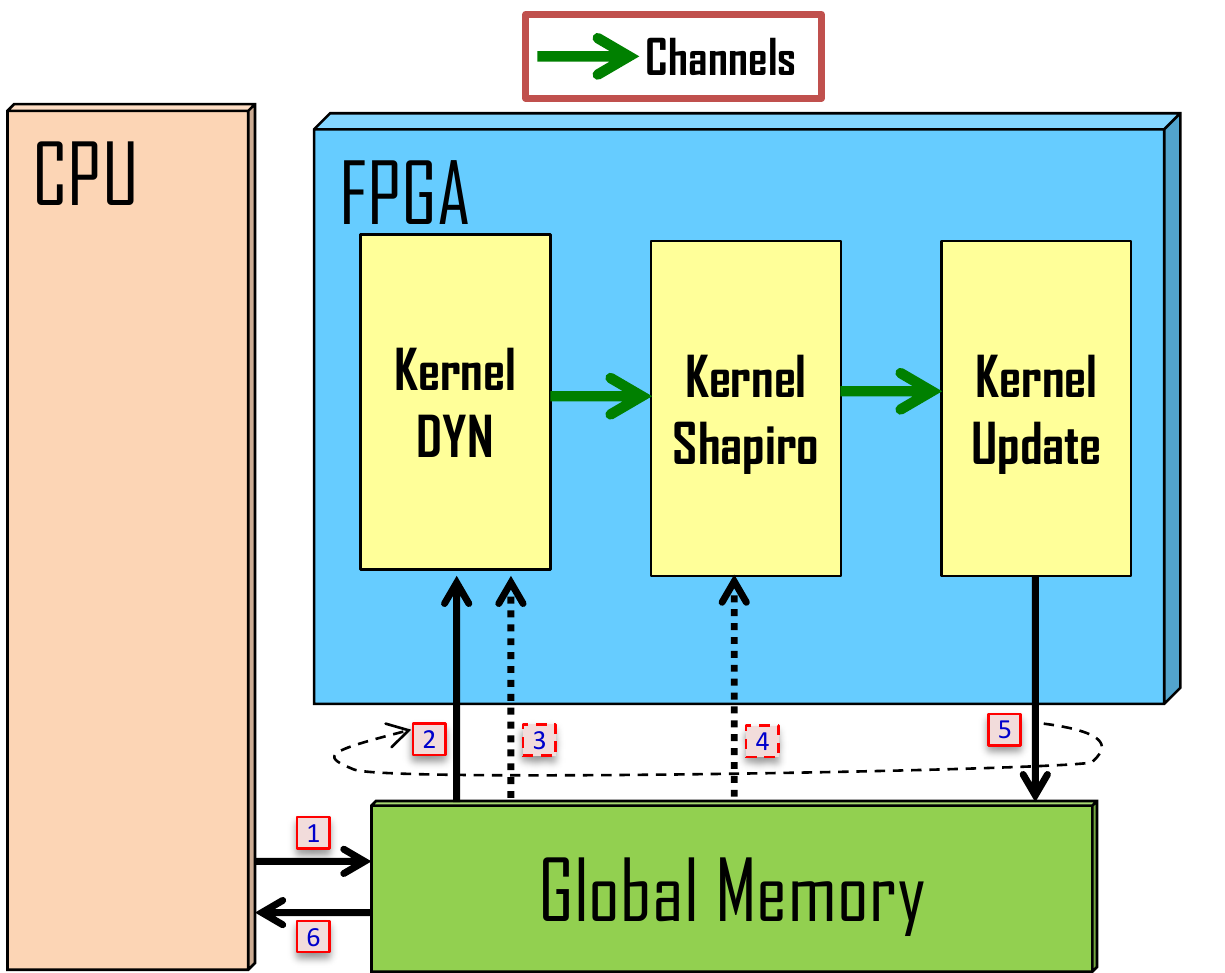}
\caption{The channelized implementation of the 2D shallow-water model on a Stratix-V FPGA using the Altera-OpenCL framework. The numbers show the sequence of operations. Note that steps 2--5 repeat $NT$ times, where $NT$ is the number of time steps of the simulation. Dotted lines indicate memory accesses required to create stencils.}
\label{fig:2dshallow_FPGA_channelized}
\end{figure}

In order to preclude the need to use the global memory for communicating between kernels, we use \emph{channels} as discussed in \autoref{sec:porting-for-perf-on-FPGAs}. The resulting architecture is shown in \autoref{fig:2dshallow_FPGA_channelized}.

The three kernels are now launched \emph{concurrently} from the host. This is achieved by creating multiple \emph{command queues} on the host using the OpenCL API. The kernels directly communicate data with each other using channels. These channel read/write calls are \emph{blocking}, which ensures the kernels are synchronized%
\footnote{Note that AOCL allows using OpenCL compliant \emph{pipes} in place of channels, but since OpenCL pipes are non-blocking, we need to add an attribute to indicate we want their blocking versions.}%
. %
While we currently introduce the channels between kernels manually, it is straightforward to update our OpenCL code generator to create these channelized kernels, as it mainly  involves replacing accesses to global memory variables with channel calls. There will be almost no change to the host code, other than launching the kernels on concurrent queues, and removing some of the arguments passed to the kernels (as most of them are no longer accessing global memory). 

\subsubsection{The Channelized Version with \emph{Smart Cache}}
Ideally, the use of channels should lead to a coarse-grained pipeline between kernels where only the first and the last kernels in the pipeline interact with the global memory. However, note from the \autoref{fig:2dshallow_FPGA_channelized} that there is additional communication between the \emph{dynamics} and the \emph{shapiro} kernel, and the global memory. This is because these kernels involve stencil computations, and they end up doing random (and redundant) accesses to the global memory to create the required stencil window. This is obviously inefficient, as the random accesses breakdown the streaming with high-latency memory accesses.

FPGAs do not have any caches, and repeated accesses to global memory, even if localized -- as would be the case for typical stencil computations -- can lead to considerable loss of performance. In order to maintain streaming in such cases, we use the available on-chip memory resources as a \emph{smart cache}, which involves using the on-chip memory in a deterministic manner to maintain a windowing buffer from the data streams. This allows us to create stencils without having to redundantly access global memories. We have also proposed some innovations for cases where we have difficult boundary conditions or very large reaches for the stencil data, but they are outside the scope of this paper. Fortunately the lateral boundary conditions in the 2d shallow-water model under consideration are \emph{closed}, and the 2- and 4-point stencils that we require for the model do not require very large reaches.

\begin{figure*}[t!]
	\centering
	\includegraphics[width=0.7\linewidth]{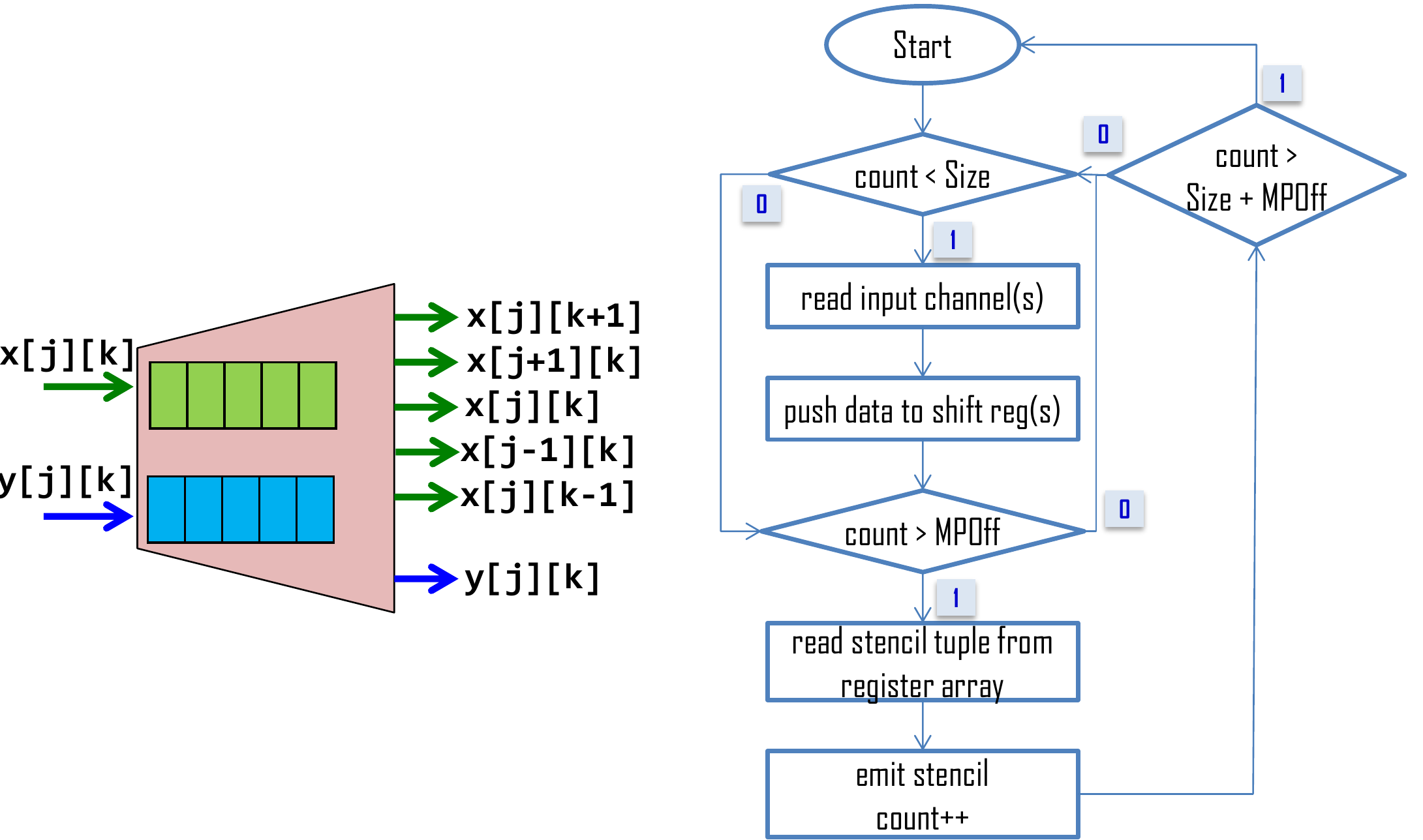}
	\caption{A high-level view of the smart cache (left) and a flowchart of its operation (right). The hypothetical case shown (left) is very similar to the requirement for 2D shallow-water model, and involves a 5-point stencil for one input stream $x$, while the other stream $y$ requires no stencil, but needs to be buffered for synchronization. The flow-chart on the right shows its implementation flow. Note: $Size$ = size of the stream, and $MPOff$ is the maximum \emph{positive} offset for a given stencil shape.}
	\label{fig:smache}
\end{figure*}

A high-level view of the smart-cache (or stencil-generator) module is shown in \autoref{fig:smache}, along with a flow-chart showing its operation. The smart-cache module can deal with stencils of arbitrary shapes, and can also synchronize multiple streams of data as shown in the figure. The flowchart is shown to highlight the fact that our OpenCL kernel for implementing this smart-cache is highly parametrizable: we simply need to set the size of the array ($Size$) and the \emph{maximum positive offset} ($MPOff$), and the loop bounds and buffer sizes are all calculated automatically. The only customization then needed in the kernel template is to create channels for the stencil tuple at the output, and assign them appropriate values from the windowing buffer.

\begin{figure*}[ht!]
	\centering
	\includegraphics[width=0.6\linewidth]{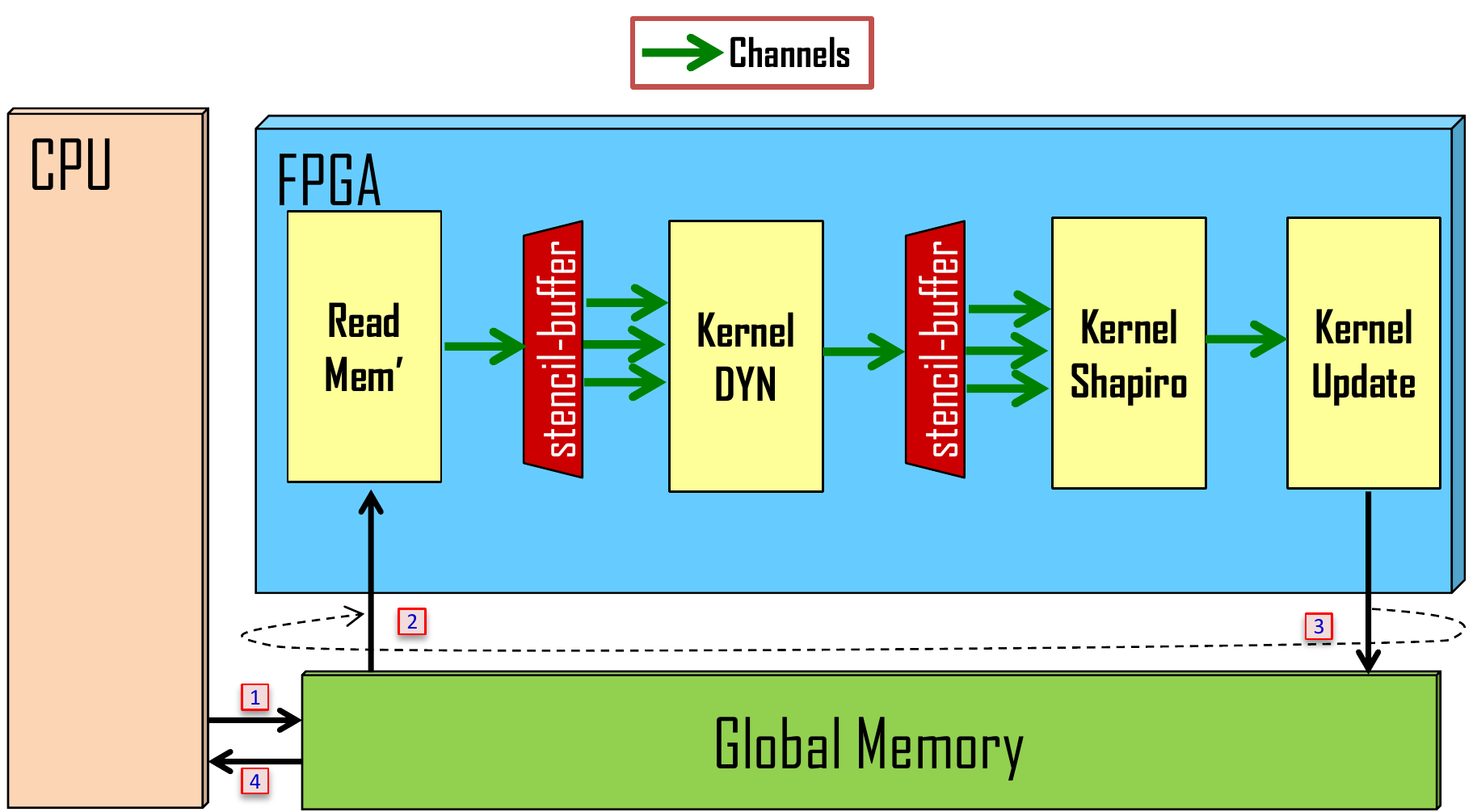}
	\caption{The channelized implementation -- along with the smart-caching module -- of the 2D shallow-water model on a Stratix-V FPGA using the Altera-OpenCL framework. The numbers show the sequence of operations. Note that steps 2--3 repeat $NT$ times, where $NT$ is the number of time steps of the simulation.}
	\label{fig:2dshallow_FPGA_channelized_smache}
\end{figure*}

When this caching module is inserted into the architecture, along with a set of kernels to read and write from the global memory, we get an efficient streaming dataflow architecture with the minimal access to global memory, as shown in \autoref{fig:2dshallow_FPGA_channelized_smache}.

\subsubsection{Experiments}

In order to evaluate the utility of our approach of creating automatic solutions for running scientific applications on FPGAs, we implemented the 2D shallow-water model on multiple targets. These targets are described in \autoref{tab:targets}:

\begin{table}
	\centering	
	\input{tables/targets.tex}
	
	\caption{Devices used for experiments.}
	\label{tab:targets}
\end{table}

\begin{figure}[t!]
	\centering
	\includegraphics[width=0.9\linewidth]{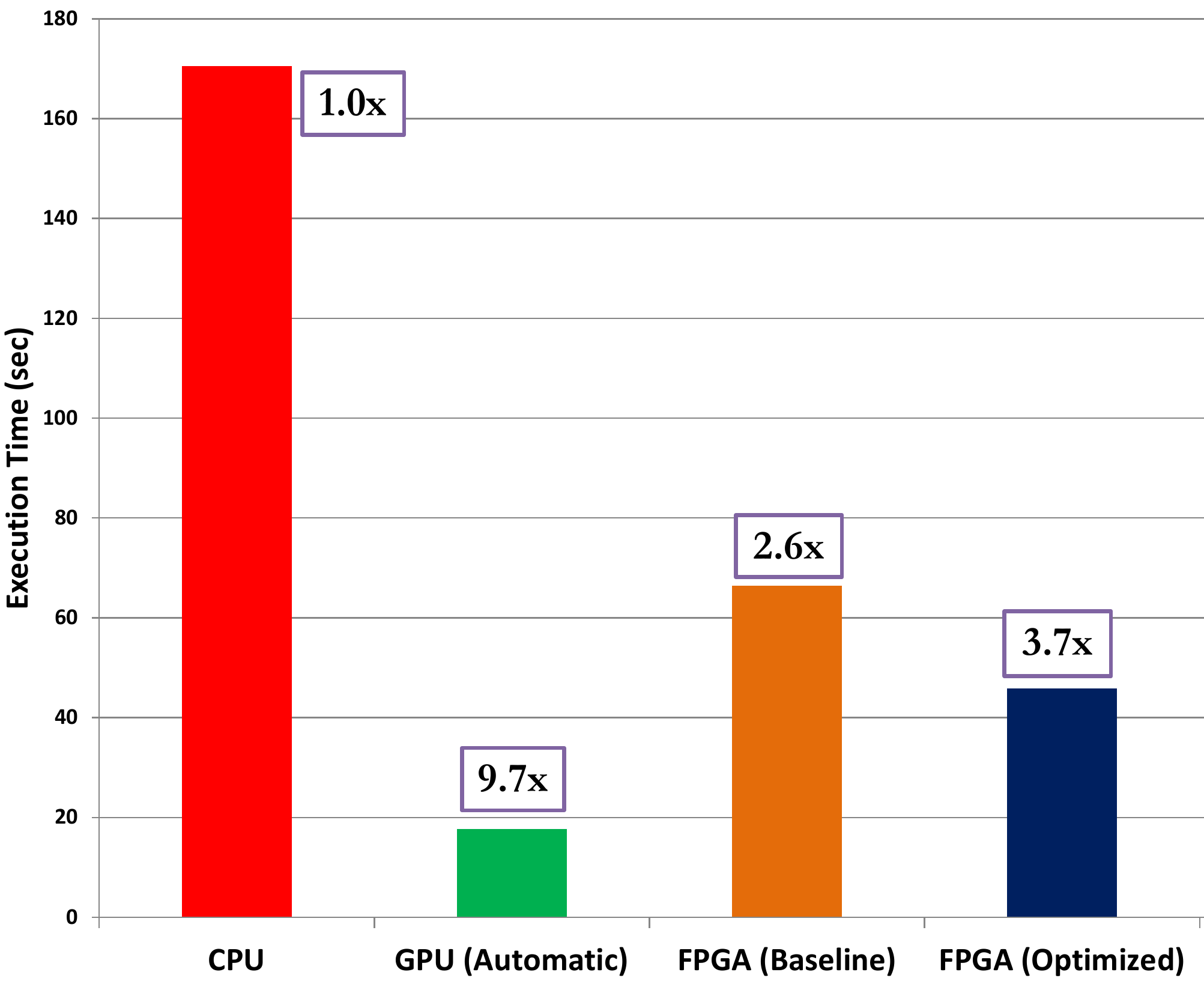}
	\caption{Execution times for running the 2D-shallow-water model on the three targets described in \autoref{tab:targets}. The model is run on a $500\times500$ grid, for $10,000$ time steps.}
	\label{fig:results-2d-shallow-waterperf}
\end{figure}

We ran the experiment for $10,000$ time steps, on a $500\times500$ grid. The execution times for the targets are shown in \autoref{fig:results-2d-shallow-waterperf}. Note that we show results for two versions of FPGA implementation, the baseline implementation, and the optimized implementation that uses channels and on-chip stencil buffers.

\begin{figure}[t!]
	\centering
	\includegraphics[width=0.9\linewidth]{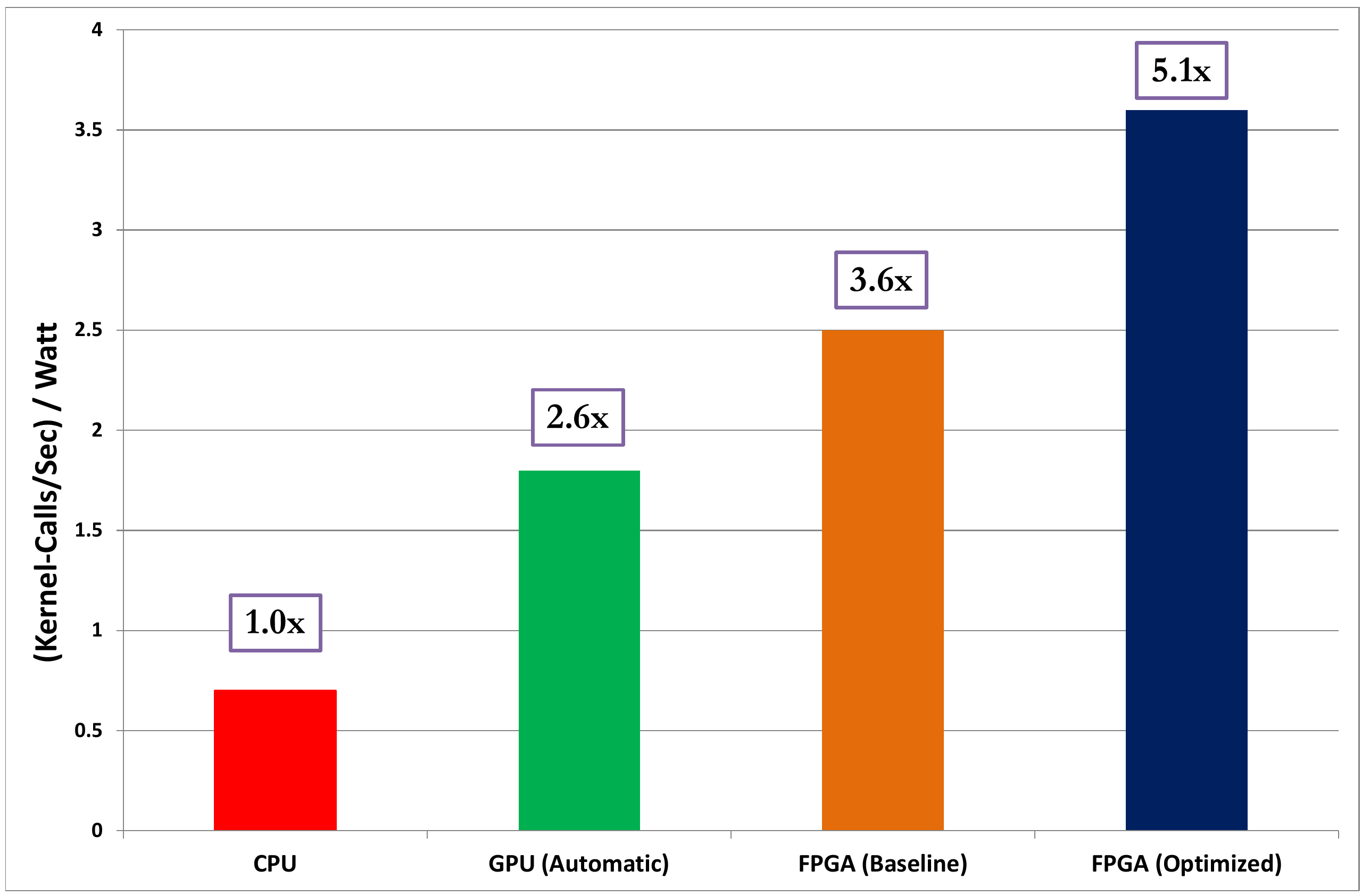}
	\caption{Energy-efficiency for running the 2D-shallow-water model on the three targets described in \autoref{tab:targets}. The model is run on a $500\times500$ grid, for $10,000$ time steps.}
	\label{fig:results-2d-shallow-water_powEff}
\end{figure}

As discussed earlier, the GPU implementation leads to an almost $10\times$ improvement, and this requires absolutely no manual coding. The baseline FPGA version uses OpenCL code that is not yet optimized, and gives $2.6\times$ performance improvement over the CPU baseline. Once we incorporate the two optimizations that we discuss in \autoref{sec:parallelize-for-fpgas}, we get a $3.7\times$ improvement over CPUs, which is a $42\%$ improvement over the baseline FPGA solution. Not surprisingly, GPUs outperform both CPU and FPGA. At the same time though, these results validate our observation that optimizations for FPGA solutions that encourage streaming can lead to significant performance improvements.

FPGAs come out a lot stronger when we take the energy-efficiency perspective. In the context of HPC (which means we are considering top-of-the-line devices), FPGAs consume around an order of magnitude less power then CPUs or GPUs. This means that FPGAs can afford to perform worse than e.g. GPUs in terms of pure throughput performance, but still be more energy-efficient, and our results confirm this, as shown in \autoref{fig:results-2d-shallow-water_powEff}. The trend in favour of FPGAs is obvious, and the optimized FPGA solution is $5.1\times$ more energy-efficient than the CPU, and $2.0\times$ more efficient than even the GPU implementation. We are already generating the complete GPU solution automatically, and our ongoing work will extend the automation to FPGA solutions, effectively enabling \emph{turn-key} solutions that lead to these kinds of performance and energy-efficiency gains.

\section{Conclusion}
Accelerator devices have firmly established themselves in the domain of HPC for scientific applications. Our work aims to make a contribution towards enabling an increasingly larger body of scientists to use these relatively low-cost devices to accelerate their applications. One of the biggest hurdles to widespread adoption of accelerators is the expertise required to program these devices for parallelism  and performance. OpenCL provides a portable, heterogeneous platform for programming such accelerators, but that still leaves a considerable body of legacy (and even new) code that is in languages like Fortran. We have developed an auto-parallelizing source-source compiler that can translate such scientific code from Fortran into parallel OpenCL code. 

In this paper we have focussed on extending this compiler to generate OpenCL code specifically for performance on FPGAs. We observed that maintaining streaming across kernels on an FPGA is critical for getting good performance, and proposed optimizations in the architecture to achieve this pattern, along with a route to extending the current compiler to enable automatic generation of optimized FPGA OpenCL code. Our results using a 2D shallow-water model as an illustration show that we are achieving up to $10\times$ better performance over CPUs by using the automatic OpenCL code on GPUs, and that with our proposed optimizations, we can create FPGA solutions which are $5\times$ and $2\times$ more energy efficient than CPU and GPU respectively, and that our proposed optimizations for FPGAs lead to a $42\%$ improvement over the baseline unoptimized FPGA solution.

The work is ongoing, and we expect to extend the Fortran-to-OpenCL compiler so that it can automatically generate the FPGA optimizations we have proposed in this paper, along with others. In the longer term we aim to extend our compiler to target a cluster of devices, which we feel is feasible within our framework.

\bibliographystyle{abbrvnat}

\bibliography{paper-HiPEAC-HLPGPU2018-oclChannels.bbl}

\end{document}

%% file: tables/targets.tex
\begin{tabular}{|c|c|}
	\hline 
	\multicolumn{2}{|c|}{Devices} \\ 
	\hline 
	CPU & Intel i3-2100 @ 3.10GHz
	\\ 
	\hline 
	GPU & GeForce GTX TITAN
	\\ 
	\hline 
	FPGA & Altera (Intel) Stratix-V
	\\ 
	\hline 
\end{tabular} 

%% file: ms.bbl
\begin{thebibliography}{19}
\providecommand{\natexlab}[1]{#1}
\providecommand{\url}[1]{\texttt{#1}}
\expandafter\ifx\csname urlstyle\endcsname\relax
  \providecommand{\doi}[1]{doi: #1}\else
  \providecommand{\doi}{doi: \begingroup \urlstyle{rm}\Url}\fi

\bibitem[Amini et~al.(2011)Amini, Ancourt, Coelho, Creusillet, Guelton,
  Irigoin, Jouvelot, Keryell, and Villalon]{amini2011pips}
M.~Amini, C.~Ancourt, F.~Coelho, B.~Creusillet, S.~Guelton, F.~Irigoin,
  P.~Jouvelot, R.~Keryell, and P.~Villalon.
\newblock Pips is not (just) polyhedral software adding gpu code generation in
  pips.
\newblock In \emph{First International Workshop on Polyhedral Compilation
  Techniques (IMPACT 2011) in conjonction with CGO 2011}, 2011.

\bibitem[Brioude et~al.(2013)Brioude, Arnold, Stohl, Cassiani, Morton, Seibert,
  Angevine, Evan, Dingwell, Fast, et~al.]{brioude2013lagrangian}
J.~Brioude, D.~Arnold, A.~Stohl, M.~Cassiani, D.~Morton, P.~Seibert,
  W.~Angevine, S.~Evan, A.~Dingwell, J.~D. Fast, et~al.
\newblock The lagrangian particle dispersion model flexpart-wrf version 3.1.
\newblock \emph{Geoscientific Model Development}, 6\penalty0 (6):\penalty0
  1889--1904, 2013.

\bibitem[Burgers et~al.(2002)Burgers, Balmaseda, Vossepoel, van Oldenborgh, and
  Van~Leeuwen]{burgers2002balanced}
G.~Burgers, M.~A. Balmaseda, F.~C. Vossepoel, G.~J. van Oldenborgh, and P.~J.
  Van~Leeuwen.
\newblock Balanced ocean-data assimilation near the equator.
\newblock \emph{Journal of physical oceanography}, 32\penalty0 (9):\penalty0
  2509--2519, 2002.

\bibitem[Czajkowski et~al.(2012)Czajkowski, Aydonat, Denisenko, Freeman,
  Kinsner, Neto, Wong, Yiannacouras, and Singh]{112.151}
T.~Czajkowski, U.~Aydonat, D.~Denisenko, J.~Freeman, M.~Kinsner, D.~Neto,
  J.~Wong, P.~Yiannacouras, and D.~Singh.
\newblock From opencl to high-performance hardware on {FPGA}s.
\newblock In \emph{Field Programmable Logic and Applications (FPL), 2012 22nd
  International Conference on}, pages 531--534, Aug 2012.
\newblock \doi{10.1109/FPL.2012.6339272}.

\bibitem[Escobar et~al.(2016)Escobar, Chang, and Valderrama]{7051218}
F.~A. Escobar, X.~Chang, and C.~Valderrama.
\newblock Suitability analysis of fpgas for heterogeneous platforms in hpc.
\newblock \emph{IEEE Transactions on Parallel and Distributed Systems},
  27\penalty0 (2):\penalty0 600--612, Feb 2016.
\newblock ISSN 1045-9219.
\newblock \doi{10.1109/TPDS.2015.2407896}.

\bibitem[K{\"a}mpf(2009)]{kampf2009ocean}
J.~K{\"a}mpf.
\newblock \emph{Ocean Modelling for Beginners: Using Open-source Software}.
\newblock Springer Science \& Business Media, 2009.

\bibitem[Liao et~al.(2010)Liao, Quinlan, Panas, and De~Supinski]{liao2010rose}
C.~Liao, D.~J. Quinlan, T.~Panas, and B.~R. De~Supinski.
\newblock A rose-based openmp 3.0 research compiler supporting multiple runtime
  libraries.
\newblock In \emph{International Workshop on OpenMP}, pages 15--28. Springer,
  2010.

\bibitem[Nabi and Vanderbauwhede(2015)]{nabi-mpstream-h2rc}
S.~W. Nabi and W.~Vanderbauwhede.
\newblock Mp-stream: A multi-platform fpga-centric memory performance
  benchmark.
\newblock In \emph{Proceedings of the Second International Workshop on
  Heterogeneous High-performance Reconfigurable Computing}, 2015.
\newblock URL \url{https://h2rc.cse.sc.edu/2016/papers/paper_18.pdf}.

\bibitem[{OpenSPL}(2017)]{openspl}
{OpenSPL}.
\newblock {The Open Spatial Programming Language: OpenSPL}, 2017.
\newblock URL \url{http://www.openspl.org/}.
\newblock At \url{http://www.openspl.org/}. Last accessed on 15th Dec 2017.

\bibitem[Orchard and Rice(2013)]{Orchard:2013:UFS:2541348.2541356}
D.~Orchard and A.~Rice.
\newblock Upgrading fortran source code using automatic refactoring.
\newblock In \emph{Proceedings of the 2013 ACM Workshop on Workshop on
  Refactoring Tools}, WRT '13, pages 29--32, New York, NY, USA, 2013. ACM.
\newblock ISBN 978-1-4503-2604-9.
\newblock \doi{10.1145/2541348.2541356}.
\newblock URL \url{http://doi.acm.org/10.1145/2541348.2541356}.

\bibitem[Overbey et~al.(2005)Overbey, Xanthos, Johnson, and
  Foote]{overbey2005refactorings}
J.~Overbey, S.~Xanthos, R.~Johnson, and B.~Foote.
\newblock Refactorings for fortran and high-performance computing.
\newblock In \emph{Proceedings of the second international workshop on Software
  engineering for high performance computing system applications}, pages
  37--39. ACM, 2005.

\bibitem[Pell and Averbukh(2012)]{112.154}
O.~Pell and V.~Averbukh.
\newblock Maximum performance computing with dataflow engines.
\newblock \emph{Computing in Science Engineering}, 14\penalty0 (4):\penalty0
  98--103, July 2012.
\newblock ISSN 1521-9615.
\newblock \doi{10.1109/MCSE.2012.78}.

\bibitem[Stone et~al.(2010)Stone, Gohara, and Shi]{stone2010opencl}
J.~E. Stone, D.~Gohara, and G.~Shi.
\newblock Opencl: A parallel programming standard for heterogeneous computing
  systems.
\newblock \emph{Computing in science \& engineering}, 12\penalty0 (3):\penalty0
  66--73, 2010.

\bibitem[Tinetti and M{\'e}ndez(2012)]{tinetti2012fortran}
F.~G. Tinetti and M.~M{\'e}ndez.
\newblock Fortran legacy software: source code update and possible
  parallelisation issues.
\newblock In \emph{ACM SIGPLAN Fortran Forum}, volume~31, pages 5--22. ACM,
  2012.

\bibitem[Vanderbauwhede and Takemi(2015)]{vanderbauwhede2015twinned}
W.~Vanderbauwhede and T.~Takemi.
\newblock {Twinned buffering: A simple and highly effective scheme for
  parallelization of Successive Over-Relaxation on GPUs and other
  accelerators}.
\newblock In \emph{High Performance Computing \& Simulation (HPCS), 2015 Int'l
  Conf on}, pages 436--443. IEEE, 2015.

\bibitem[Watanabe and Jin(2003)]{watanabe2003moist}
M.~Watanabe and F.-F. Jin.
\newblock A moist linear baroclinic model: Coupled dynamical--convective
  response to el ni{\~n}o.
\newblock \emph{Journal of climate}, 16\penalty0 (8):\penalty0 1121--1139,
  2003.

\bibitem[Weller et~al.(2017)Weller, Oboril, Lukarski, Becker, and
  Tahoori]{Weller:2017:EES:3020078.3021730}
D.~Weller, F.~Oboril, D.~Lukarski, J.~Becker, and M.~Tahoori.
\newblock Energy efficient scientific computing on fpgas using opencl.
\newblock In \emph{Proceedings of the 2017 ACM/SIGDA International Symposium on
  Field-Programmable Gate Arrays}, FPGA '17, pages 247--256, New York, NY, USA,
  2017. ACM.
\newblock ISBN 978-1-4503-4354-1.
\newblock \doi{10.1145/3020078.3021730}.
\newblock URL \url{http://doi.acm.org/10.1145/3020078.3021730}.

\bibitem[{Xilinx}(2014{\natexlab{a}})]{sdaccel}
{Xilinx}.
\newblock {The Xilinx SDAccel Development Environment}, 2014{\natexlab{a}}.
\newblock URL
  \url{http://www.xilinx.com/publications/prod_mktg/sdx/sdaccel-backgrounder.pdf}.

\bibitem[{Xilinx}(2014{\natexlab{b}})]{vivado}
{Xilinx}.
\newblock {Vivado HLx Editions; Bringing Ultra High Productivity to Mainstream
  Systems \& Platform Designers}, 2014{\natexlab{b}}.
\newblock URL
  \url{https://www.xilinx.com/support/documentation/backgrounders/vivado-hlx.pdf}.

\end{thebibliography}
